\documentclass[10pt,conference,letter]{IEEEtran}
\usepackage{graphicx}
\usepackage[cmex10,tbtags]{amsmath} 
\usepackage{booktabs}
\usepackage{cite}
\usepackage{epic,eepic}
\usepackage{hyperref}
\usepackage{fancybox,framed}
\newtheorem{thm}{Theorem}
\IEEEoverridecommandlockouts
\begin{document}

\title{Matrix Adaptive Synthesis Filter for Uniform Filter Bank}

\author{
\IEEEauthorblockN{Sandeep Patel, Ravindra Dhuli and Brejesh Lall}
\IEEEauthorblockA{Department of Electrical Engineering\\
Indian Institute of Technology Delhi \\
New Delhi, India -110016\\
Email: sanpatel90@gmail.com, ravindra\_dhuli@yahoo.co.in, brejesh@ee.iitd.ac.in} 
}

\doublebox{\begin{minipage}{1\textwidth}
\begin{center}
\textbf{This is the accepted version of the paper}
\end{center}
\medskip
Link to IEEE version: \href{https://ieeexplore.ieee.org/abstract/document/6487978}{https://ieeexplore.ieee.org/abstract/document/6487978} 

\medskip

\textbf{Digital Object Identifier 10.1109/NCC.2013.6487978}

\medskip

Citation: S. Patel, R. Dhuli and B. Lall, "Matrix adaptive synthesis filter for uniform filter bank," 2013 National Conference on Communications (NCC), 2013, pp. 1-5, doi: 10.1109/NCC.2013.6487978.

\medskip

\textcopyright 2013 IEEE.  Personal use of this material is permitted.  Permission from IEEE must be obtained for all other uses, in any current or future media, including reprinting/republishing this material for advertising or promotional purposes, creating new collective works, for resale or redistribution to servers or lists, or reuse of any copyrighted component of this work in other works.
\end{minipage}}

\clearpage
\maketitle

\begin{abstract}
In this paper, we use a matrix adaptive filter as the synthesis stage of a Uniform Filter Bank (UFB) to reconstruct the input signal. We first develop the mathematical theory behind it by applying the model of optimal filtering at the synthesis stage of the UFB and obtaining an expression for the matrix Wiener filter. We have developed a theorem which we use to simplify the expression further. In the absence of required information about the analysis stage, we use adaptive filtering to arrive at the Wiener solution. We use the Least Mean Square (LMS) algorithm to update the filter coefficients. Through experimental results, we find that the adaptive filter is convergent for a stable Wiener filter.

\end{abstract}

\section{Introduction}
Multirate signal processing finds application in diverse areas like filter design, sub-band coding, communication etc \cite{bk:vaidyn}. Its advantage lies in reduced data rate and reduced computational cost. Filter banks play a crucial role in multirate systems. A Uniform Filter Bank (UFB) is a filter bank  with same decimation factor for each channel. If the number of channels are same as the decimation factor, then it is called a maximally decimated UFB \cite{phd:multirate}. Perfect reconstruction is a desired property for many applications and it is an active area of research \cite{bk:vaidyn}. Consider a situation where we observe a phenomenon through a number of channels, and each channel is selectively looking at a range of frequencies. To reduce data rate, we can decimate each channel output. This whole process can be modeled as the analysis stage of a UFB. If the analysis bank is completely known, we can use the matrix Wiener filter to reconstruct the input. In the absence of sufficient knowledge about the analysis bank system functions, adaptive filtering provides a mechanism to recover back the input. This is what we explore in this paper.

V. P. Sathe et al. \cite{jn:sathe} used a matrix adaptive filter for the identification of band-limited channels and compared its performance to that of a scalar adaptive filter. We, on the other hand, use it for reconstruction of the input. P. P. Vaidyanathan et al.\cite{conf:chen} used a matrix Wiener filter in the presence of sub-band quantizers and expressed it in terms of joint statistics of appropriate signals. We do not consider any quantizer. We start from a Wiener filter expression which is in terms of spectral densities of the input and desired signals, and significantly simplify it further.

\subsection{Outline}
In this paper, Section-II contains a derivation of the matrix Wiener filter for a UFB. Section-III then discusses matrix adaptive filtering for the Wiener solution. In section-IV, we provide the experimental results. Section-V, finally, contains the conclusion. 

\subsection{Notation}
All vector or matrix quantities are denoted by boldface. $\mathbf{A}^\dagger(z)$ denotes conjugate-transpose of $\mathbf{A}(z)$. $\mathbf{A}_{*}(z)$ is used to denote conjugation of coefficients of $z$ in $\mathbf{A}(z)$. $\tilde{\mathbf{A}}(z)$ denotes $\mathbf{A}_{*}^T(z^{-1})$ \cite{bk:vaidyn}.

\section{Matrix Wiener Filter for UFB}
\begin{figure}[!t]
\centering
\includegraphics[width=0.8\columnwidth,keepaspectratio=true]{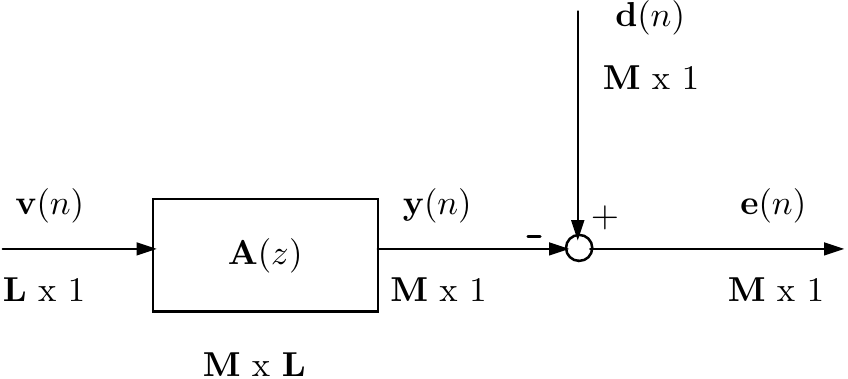}
\caption{Optimum filtering block diagram}
\label{fig_op}
\vspace{-12pt}
\end{figure}
\begin{figure*}[!t]
\centering
\includegraphics[scale=0.7]{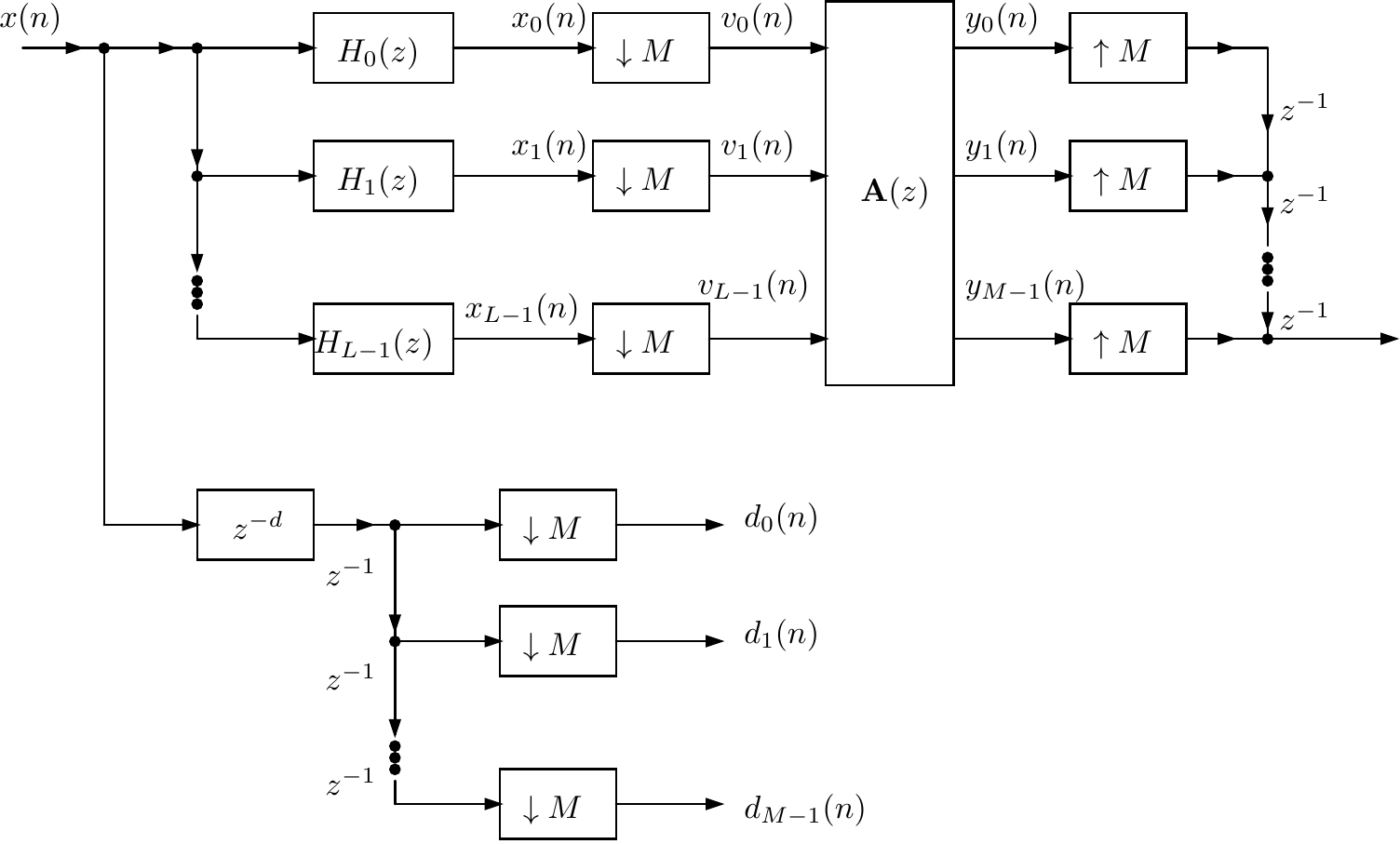}
\caption{A uniform filter bank with a Wiener or an adaptive filter based synthesis stage}
\label{fig_mb}
\vspace{-12pt}
\end{figure*}
The Wiener filter is a linear optimum filter that minimizes the mean-square error in a signal. It requires joint stationarity of the input signal and the desired signal. In case of multiple-input multiple-output systems, such type of filters are known as matrix Wiener filters. Fig.~\ref{fig_op} shows the block diagram of a matrix Wiener filter. The signal $\mathbf{v}(n)$ is the input to the filter and $\mathbf{d}(n)$ is the desired signal. The filter $\mathbf{A}(z)$ is of the following form:
\begin{equation}
\mathbf{A}(z) = \begin{bmatrix}
A_{0,0}(z) & A_{0,1}(z) & \dots & A_{0,L-1}(z) \\
A_{1,0}(z) & A_{1,1}(z) & \dots & A_{1,L-1}(z) \\
\vdots & \vdots & \ddots & \vdots \\
A_{M-1,0}(z) & A_{M-1,1}(z) & \dots & A_{M-1,L-1}(z) 
\end{bmatrix}.
\end{equation}
This form utilizes all the cross-correlations between the components of the input signal. It is more general than a set of $L$ scalar filters. If the scalar filters give an optimal solution, then we will automatically obtain a diagonal matrix as the Wiener solution. The Wiener solution of the filter can easily be derived \cite{jn:sathe} and is given by
\begin{equation}
\mathbf{A}(z) = \tilde{\mathbf{S}}_{vd}(z) \mathbf{S}^{-1}_{vv}(z)
\label{eq_w1}
\end{equation}
where $\mathbf{S}_{vv}(z)$ is the Power Spectral Density ~(PSD) of the input signal, and $\mathbf{S}_{vd}(z)$ is Cross-Spectral Density~(CSD) between the input signal and the desired signal. As $ \tilde{\mathbf{S}}_{vd}(z) = \mathbf{S}_{dv}(z)$, (\ref{eq_w1}) can be written as
\begin{equation}
\label{eq_wiener}
\mathbf{A}(z) = \mathbf{S}_{dv}(z) \mathbf{S}^{-1}_{vv}(z).
\end{equation}

A UFB structure which uses a Wiener or an adaptive filter is shown in Fig. \ref{fig_mb}. The analysis stage output $\mathbf{v}(n)$ acts as the input to the filter $\mathbf{A}(z)$. The desired signal $\mathbf{d}(n)$ is generated from $x(n)$ after delaying it by $d$ samples. An appropriate delay is needed to ensure causality of the Wiener filter. If $x(n)$ is wide sense stationary~(WSS), then $\mathbf{x}(n)$, $\mathbf{v}(n)$, and $\mathbf{d}(n)$ are also WSS~\cite{jn:brejeshsir,jn:sathe}. The cross-correlation between $\mathbf{d}(n)$ and $\mathbf{v}(n)$ is given by
\begin{align}
\medmuskip=1mu
[\mathbf{R}_{dv}(n,k)]_{i,j} 
&= \mathbf{E}[d_i(n)v_j^*(n-k)] \notag \\
&= \sum_l h_j^*(l) R_{xx}(Mk + l - i - d).
\label{eq_rdv}
\end{align}
Thus, the signals $\mathbf{d}(n)$ and $\mathbf{v}(n)$ are jointly WSS, and we can apply Wiener filtering in this setting.

We now simplify the expression for the matrix Wiener filter in the context of UFBs. Consider the system shown in Fig.~\ref{fig_mb}. The power spectral density of the signal $\mathbf{x}(n)$ is~\cite{jn:sathe}
\begin{equation}
\mathbf{S}_{xx}(z) = \mathbf{H}(z)S_{xx}(z)\mathbf{\tilde{H}}(z)
\end{equation}
where
\begin{equation}
\mathbf{H}(z) = [ H_0(z),H_1(z),\dots,H_{L-1}(z)]^T  .
\end{equation}
Therefore, the power spectral density of $\mathbf{v}(n)$ is
\begin{equation}
\mathbf{S}_{vv}(z)=(\mathbf{H}(z)S_{xx}(z)\mathbf{\tilde{H}}(z))_{\downarrow M} .
\label{eq_svv}
\end{equation}
Taking $z$-transform of~(\ref{eq_rdv}), we obtain
\begin{align}
\left[\mathbf{S}_{dv}(z)\right]_{i,j} &= \frac{1}{M} \sum_{q=0}^{M-1} (z^{\frac{1}{M}}W_M^q)^{-(i+d)} \tilde{H}_j(z^{\frac{1}{M}}W_M^q)  \notag \\
& \qquad S_{xx}(z^{\frac{1}{M}}W_M^q) \notag \\
&= ( z^{-(d+i)}S_{xx}(z)\tilde{H}_j(z))_{\downarrow M} , \notag \\
\mathbf{S}_{dv}(z) &= ( z^{-d}\mathbf{e}_M(z) S_{xx}(z) \mathbf{\tilde{H}}(z))_{\downarrow M} 
\label{eq_svd}
\end{align}
where $\mathbf{e}_M(z)$ is the delay chain vector given by 
\begin{equation}
\mathbf{e}_M(z) = [1, z^{-1}, \dots, z^{-(M-1)}]^T .
\end{equation}
Using (\ref{eq_wiener}),(\ref{eq_svv}) and (\ref{eq_svd}), we obtain
\begin{equation}
\resizebox{1\linewidth}{!}{$
\mathbf{A}(z) = \left( z^{-d}\mathbf{e}_M(z) S_{xx}(z) \mathbf{\tilde{H}}(z)\right)_{\downarrow M} \left(\left(\mathbf{H}(z)S_{xx}(z)\mathbf{\tilde{H}}(z)\right)_{\downarrow M} \right)^{-1} .$}
\end{equation} 
Further simplification is not obvious. To simplify it further, we have developed some tools which we now present.

\subsection{Determinant of submatrix of PSD matrix $\mathbf{S}_{vv}(z)$ }
We now state a theorem. Any $Q\times Q$ submatrix of $\mathbf{S}_{vv}(z)$, which is obtained by choosing any $Q$ rows and any $Q$ columns, is given  by
\begin{equation}
\arraycolsep=2pt
\resizebox{1\columnwidth}{!}{$
\begin{bmatrix}
(H_{r_1}\tilde{H}_{c_1} S_{xx})_{\downarrow M} & (H_{r_1}\tilde{H}_{c_2} S_{xx})_{\downarrow M} & \dots &  (H_{r_1}\tilde{H}_{c_Q} S_{xx})_{\downarrow M} \\
(H_{r_2}\tilde{H}_{c_1} S_{xx})_{\downarrow M} & (H_{r_2}\tilde{H}_{c_2} S_{xx})_{\downarrow M} & \dots &  (H_{r_2}\tilde{H}_{c_Q} S_{xx})_{\downarrow M} \\
\vdots & \vdots & \ddots & \vdots \\ 
(H_{r_Q}\tilde{H}_{c_1} S_{xx})_{\downarrow M} & (H_{r_Q}\tilde{H}_{c_2} S_{xx})_{\downarrow M} & \dots &  (H_{r_Q}\tilde{H}_{c_Q} S_{xx})_{\downarrow M}
\end{bmatrix} .$}
\label{eq_sub-block}
\end{equation}
\begin{thm}
The determinant of the submatrix (\ref{eq_sub-block}) is given by
\begin{equation}
\begin{split}
&\Delta_M(Q) \\
&= \frac{1}{M^Q} \sum_{i_1 = 0}^{M-Q} \sum_{i_2 > i_1}^{M-Q+1} \dots \sum_{i_Q > i_{Q-1}}^{M-1} S_{xx}(z^{\frac{1}{M}}W^{i_1})S_{xx}(z^{\frac{1}{M}}W^{i_2}) \\
& \quad \dotsm S_{xx}(z^{\frac{1}{M}}W^{i_Q}) E_{H_{r_1},H_{r_2},\dots,H_{r_Q}}(i_1,i_2,\dots,i_Q) \\ 
& \quad \tilde{E}_{H_{c_1},H_{c_2},\dots,H_{c_Q}}(i_1,i_2,\dots,i_Q) 
\label{eq_det}
\end{split}
\end{equation}
where
\begin{equation}
\begin{split}
&E_{H_{r_1},H_{r_2},\dots,H_{r_Q}}(i_1,i_2,\dots,i_Q) \\
=& \begin{vmatrix}
H_{r_1}(z^{\frac{1}{M}}W^{i_1}) & H_{r_1}(z^{\frac{1}{M}}W^{i_2}) & \dots & H_{r_1}(z^{\frac{1}{M}}W^{i_Q}) \\
H_{r_2}(z^{\frac{1}{M}}W^{i_1}) & H_{r_2}(z^{\frac{1}{M}}W^{i_2}) & \dots & H_{r_2}(z^{\frac{1}{M}}W^{i_Q}) \\
\vdots & \vdots & \ddots & \vdots \\ 
H_{r_Q}(z^{\frac{1}{M}}W^{i_1}) & H_{r_Q}(z^{\frac{1}{M}}W^{i_2}) & \dots & H_{r_Q}(z^{\frac{1}{M}}W^{i_Q})
\end{vmatrix} .
\end{split}
\label{eq_edet}
\end{equation}
\end{thm}
\begin{IEEEproof}
We use mathematical induction to prove the theorem. For $Q = 1$, we can easily verify that it is true. Assuming it to be true for $Q$, we have to prove it true for $Q+1$. By expanding the determinant along first row, we obtain
\begin{equation}
\Delta_M(Q+1) = \sum_{p=1}^{Q+1} (H_{r_1}\tilde{H}_{c_p} S_{xx})_{\downarrow M} C_{1,p}
\end{equation} 
where $C_{1,p}$ is the co-factor of the element $(1,p)$.
\begin{align}
&\Delta_M(Q+1) \notag\\
& = \frac{1}{M^Q} \sum_{i_1 = 0}^{M-Q} \sum_{i_2 > i_1}^{M-Q+1} \dots \sum_{i_Q > i_{Q-1}}^{M-1} S_{xx}(W^{i_1}) S_{xx}(W^{i_2}) \dotsm \notag\\
& S_{xx}(W^{i_Q}) E_{H_{r_2},H_{r_3},\dots,H_{r_{Q+1}}}(i_1,i_2,\dots,i_Q) \sum_{p=1}^{Q+1} (-1)^{p+1} \notag\\ 
&(H_{r_1}\tilde{H}_{c_p} S_{xx})_{\downarrow M} \tilde{E}_{H_{c_1},\dots,H_{c_{p-1}},H_{c_{p+1}},\dots,H_{c_{Q+1}}}(i_1,i_2,\dots,i_Q) \notag\\
& = K \sum_{p=1}^{Q+1} (-1)^{p+1} \frac{1}{M} \sum_{q=0}^{M-1} H_{r_1}(W^q)\tilde{H}_{c_p}(W^q) S_{xx}(W^q)\notag\\ 
& \tilde{E}_{H_{c_1},\dots,H_{c_{p-1}},H_{c_{p+1}},\dots,H_{c_{Q+1}}}(i_1,i_2,\dots,i_Q) \notag\\
&= \frac{K}{M} \sum_{q=0}^{M-1} H_{r_1}(W^q) S_{xx}(W^q) \sum_{p=1}^{Q+1} \tilde{H}_{c_p}(W^q) (-1)^{p+1} \notag\\ 
& \tilde{E}_{H_{c_1},\dots,H_{c_{p-1}},H_{c_{p+1}},\dots,H_{c_{Q+1}}}(i_1,i_2,\dots,i_Q) \notag\\
&= \frac{K}{M} \sum_{q=0}^{M-1} H_{r_1}(W^q) S_{xx}(W^q) \notag\\ 
& \tilde{E}_{H_{c_1},H_{c_2},\dots,H_{c_{Q+1}}}(q,i_1,i_2,\dots,i_Q) .
\end{align}
If $q$ is equal to any of $i_1,i_2,\dots,i_Q$, then two rows of the determinant $\tilde{E}_{H_{c_1},H_{c_2},\dots,H_{c_{Q+1}}}(q,i_1,i_2,\dots,i_Q)$ will be same and its value will be zero.
\begin{align}
&\Delta_M(Q+1) = \frac{1}{M^{Q+1}} \sum_{i_1 = 0}^{M-Q} \sum_{i_2 > i_1}^{M-Q+1} \dots \sum_{i_Q > i_{Q-1}}^{M-1} \sum_{q \neq i_1,i_2,\dots,i_Q} \notag \\
& \qquad S_{xx}(W^{i_1}) S_{xx}(W^{i_2}) \dotsm S_{xx}(W^{i_Q}) S_{xx}(W^q) H_{r_1}(W^q) \notag \\
& \qquad E_{H_{r_2},H_{r_3},\dots,H_{r_{Q+1}}}(i_1,i_2,\dots,i_Q) \notag \\ 
& \qquad \tilde{E}_{H_{c_1},H_{c_2},\dots,H_{c_{Q+1}}}(q,i_1,i_2,\dots,i_Q) .
\end{align}
Adding up all the terms with the same combination $S_{xx}(W^{j_1}) \allowbreak S_{xx}(W^{j_2}) \dotsm S_{xx}(W^{j_{Q+1}})$ where $j_1 < j_2 <\dots <j_{Q+1}$, we obtain
\begin{align}
&\Delta_M(Q+1) \notag \\
&=\frac{1}{M^{Q+1}} \sum_{j_1 = 0}^{M-Q-1} \sum_{j_2 > j_1}^{M-Q} \dots \sum_{j_{Q+1} > j_Q}^{M-1} S_{xx}(W^{j_1})S_{xx}(W^{j_2}) \dotsm \notag \\
& S_{xx}(W^{j_{Q+1}}) \Big[ \sum_{p=1}^{Q+1} H_{r_1}(W^{j_p}) \notag \\
& E_{H_{r_2},H_{r_3},\dots,H_{r_{Q+1}}}(j_1,\dots,j_{p-1},j_{p+1},\dots,j_{Q+1}) \notag \\
& \tilde{E}_{H_{c_1},H_{c_2},\dots,H_{c_{Q+1}}}(j_p,j_1,\dots,j_{p-1},j_{p+1}, \dots, j_{Q+1}) \Big] \notag \\
&= K' \tilde{E}_{H_{c_1},H_{c_2},\dots,H_{c_{Q+1}}}(j_1,j_2,\dots,j_{Q+1}) \Big[\sum_{p=1}^{Q+1}(-1)^{p-1} \notag \\
& H_{r_1}(W^{j_p}) E_{H_{r_2},H_{r_3},\dots,H_{r_{Q+1}}}(j_1,\dots,j_{p-1},j_{p+1},\dots,j_{Q+1}) \Big] \notag \\
&= K' \tilde{E}_{H_{c_1},H_{c_2},\dots,H_{c_{Q+1}}}(j_1,j_2,\dots,j_{Q+1}) \notag \\
& E_{H_{r_1},H_{r_2},\dots,H_{r_{Q+1}}}(j_1, j_2,\dots, j_{Q+1}) \notag \\
&= \frac{1}{M^{Q+1}} \sum_{j_1 = 0}^{M-Q-1} \sum_{j_2 > j_1}^{M-Q} \dots \sum_{j_{Q+1} > j_Q}^{M-1} S_{xx}(W^{j_1})S_{xx}(W^{j_2}) \dotsm \notag \\
& S_{xx}(W^{j_{Q+1}}) E_{H_{r_1},H_{r_2},\dots,H_{r_{Q+1}}}(j_1, j_2,\dots, j_{Q+1}) \notag \\
& \tilde{E}_{H_{c_1},H_{c_2},\dots,H_{c_{Q+1}}}(j_1,j_2,\dots,j_{Q+1})
\end{align}
which proves that the statement is true for $Q+1$. We have proved it for an arbitrary value of $M$. Hence, by induction, it is true for any $M$ and $Q$.
\end{IEEEproof}

\subsection{Obtaining the Wiener filter $\mathbf{A}(z)$}
The inverse of $\mathbf{S}_{vv}(z)$ is given by
\begin{equation}
\mathbf{S}_{vv}^{-1}(z) = \frac{1}{\Delta} \begin{bmatrix}
C_{0,0} & C_{1,0} & \dots & C_{L-1,0} \\
C_{0,1} & C_{1,1} & \dots & C_{L-1,1} \\
\vdots & \vdots & \ddots & \vdots \\ 
C_{0,L-1} & C_{1,L-1} & \dots & C_{L-1,L-1}
\end{bmatrix}
\label{eq_svvin}
\end{equation}
where $\Delta$ is the determinant and $C_{i,j}$ are the co-factors of $\mathbf{S}_{vv}(z)$. Using (\ref{eq_wiener}), we can write
\begin{equation}
A_{i,j}(z) = \sum_{k=0}^{L-1}[\mathbf{S}_{dv}(z)]_{i,k} [\mathbf{S}^{-1}_{vv}(z)]_{k,j} .
\label{eq_aijtmp1}
\end{equation}
We can simplify~(\ref{eq_aijtmp1}) using (\ref{eq_svd}), (\ref{eq_svvin}) and (\ref{eq_det}).
\begin{align}
&A_{i,j}(z) \notag\\
&= \frac{1}{\Delta} \sum_{k=0}^{L-1}(z^{-(d+i)} \tilde{H}_k(z) S_{xx}(z))_{\downarrow M} C_{j,k} \notag\\
&= \frac{1}{\Delta} \frac{1}{M^{L-1}} \sum_{i_1 = 0}^{M-L+1} \sum_{i_2 > i_1}^{M-L+2} \dots \sum_{i_{L-1} > i_{L-2} }^{M-1} S_{xx}(W^{i_1}) \notag\\
& \dotsm S_{xx}(W^{i_{L-1}}) E_{H_0,\dots, H_{j-1},H_{j+1},\dots, H_{L-1}}(i_1,i_2,\dots, i_{L-1}) \notag\\
& \sum_{k=0}^{L-1}\frac{1}{M}\sum_{p=0}^{M-1} z^{-\frac{d+i}{M}} W^{-p(d+i)}\tilde{H}_k(W^p) S_{xx}(W^p)(-1)^{j+k} \notag\\
& \tilde{E}_{H_0,\dots, H_{k-1},H_{k+1},\dots, H_{L-1}}(i_1,i_2,\dots, i_{L-1}) \notag\\
&= K_1 \sum_{p=0}^{M-1} W^{-p(d+i)} S_{xx}(W^p) \sum_{k=0}^{L-1} (-1)^k \tilde{H}_k(W^p) \notag \\
& \tilde{E}_{H_0,\dots, H_{k-1},H_{k+1},\dots, H_{L-1}}(i_1,i_2,\dots, i_{L-1}) \notag \\
&= K_1 \sum_{p=0}^{M-1} W^{-p(d+i)} S_{xx}(W^p) \notag \\
& \tilde{E}_{H_0,H_1,\dots, H_{L-1}}(p,i_1,i_2,\dots, i_{L-1}) \notag \\
&= \frac{1}{\Delta} \frac{1}{M^L} \sum_{i_1 = 0}^{M-L+1} \sum_{i_2 > i_1}^{M-L+2} \dots \sum_{i_{L-1} > i_{L-2} }^{M-1} \sum_{p \neq i_1,i_2,\dots, i_{L-1}} \notag \\
& S_{xx}(W^{i_1}) S_{xx}(W^{i_2}) \dotsm S_{xx}(W^{i_{L-1}})S_{xx}(W^p) \notag \\
& E_{H_0,\dots, H_{j-1},H_{j+1},\dots, H_{L-1}}(i_1,i_2,\dots, i_{L-1})(-1)^j z^{-\frac{d+i}{M}} \notag \\
& W^{-p(d+i)} \tilde{E}_{H_0,H_1,\dots, H_{L-1}}(p,i_1,i_2,\dots, i_{L-1}) .
\end{align}
Adding up all the terms with the same combination $S_{xx}(W^{j_1})\allowbreak S_{xx}(W^{j_2}) \dotsm S_{xx}(W^{j_L})$ where $j_1 < j_2 <\dots <j_L$, we obtain
\begin{align}
&A_{i,j}(z) \notag\\
&= \frac{1}{\Delta} \frac{1}{M^L} \sum_{j_1 = 0}^{M-L} \sum_{j_2 > j_1}^{M-L+1} \dots \sum_{j_L > j_{L-1} }^{M-1} S_{xx}(W^{j_1}) S_{xx}(W^{j_2}) \dotsm \notag\\
& S_{xx}(W^{j_L}) (-1)^j z^{-\frac{d+i}{M}} \sum_{q=1}^L W^{-j_q(d+i)} \notag\\
& E_{H_0,\dots, H_{j-1},H_{j+1},\dots, H_{L-1}}(j_1,\dots,j_{q-1},j_{q+1},\dots,j_L) \notag\\
& \tilde{E}_{H_0,H_1,\dots, H_{L-1}}(j_q,j_1,\dots,j_{q-1},j_{q+1},\dots,j_L ) \notag\\
&= K_2 \tilde{E}_{H_0,\dots, H_{L-1}}(j_1,j_2,\dots,j_L ) \sum_{q=1}^L (-1)^{j+q-1} W^{-j_q(d+i)} \notag \\
& z^{-\frac{d+i}{M}}E_{H_0,\dots, H_{j-1},H_{j+1},\dots, H_{L-1}}(j_1,\dots,j_{q-1},j_{q+1},\dots,j_L) \notag \\
&= K_2 \tilde{E}_{H_0,H_1,\dots, H_{L-1}}(j_1,j_2,\dots,j_L ) \notag \\
& E_{H_0,\dots, H_{j-1},z^{-(d+i)},H_{j+1},\dots, H_{L-1}}(j_1,j_2,\dots,j_L) \notag \\
&= \frac{1}{\Delta} \frac{1}{M^L} \sum_{j_1 = 0}^{M-L} \sum_{j_2 > j_1}^{M-L+1} \dots \sum_{j_L > j_{L-1} }^{M-1} S_{xx}(W^{j_1}) S_{xx}(W^{j_2}) \dotsm \notag \\
& S_{xx}(W^{j_L}) \tilde{E}_{H_0,H_1,\dots, H_{L-1}}(j_1,j_2,\dots,j_L ) \notag \\
& E_{H_0,\dots, H_{j-1},z^{-(d+i)},H_{j+1},\dots, H_{L-1}}(j_1,j_2,\dots,j_L) .
\label{eq_aij}
\end{align} 
From (\ref{eq_aij}), it is clear that $\mathbf{A}(z)$, in general, is dependent on the PSD $S_{xx}(z)$ unless $L = M$. Thus, the UFB has to be maximally decimated for the Wiener filter to be independent of the PSD. For the maximally decimated case, $\mathbf{A}(z)$ reduces to 
\begin{equation}
\resizebox{1\columnwidth}{!}{$
A_{i,j}(z) = \frac{E_{H_0,\dots, H_{j-1},z^{-(d+i)},H_{j+1},\dots, H_{M-1}}(0,1,\dots,M-1)}{E_{H_0,H_1,\dots, H_{M-1}}(0,1,\dots,M-1)} .$}
\end{equation}
Using (\ref{eq_edet}), we can write
\begin{equation}
\resizebox{1\columnwidth}{!}{$
\arraycolsep = 2pt
A_{i,j}(z) = \frac{\begin{vmatrix}
H_0(z^{\frac{1}{M}}) & H_0(z^{\frac{1}{M}}W) & \dots & H_0(z^{\frac{1}{M}}W^{M-1}) \\
\vdots & \vdots & \ddots & \vdots \\
H_{j-1}(z^{\frac{1}{M}}) & H_{j-1}(z^{\frac{1}{M}}W) & \dots & H_{j-1}(z^{\frac{1}{M}}W^{M-1}) \\
z^{-\frac{d+i}{M}} & (z^{\frac{1}{M}}W)^{-(d+i)} & \dots & (z^{\frac{1}{M}}W^{M-1})^{-(d+i)} \\
H_{j+1}(z^{\frac{1}{M}}) & H_{j+1}(z^{\frac{1}{M}}W) & \dots & H_{j+1}(z^{\frac{1}{M}}W^{M-1}) \\
\vdots & \vdots & \ddots & \vdots \\ 
H_{M-1}(z^{\frac{1}{M}}) & H_{M-1}(z^{\frac{1}{M}}W) & \dots & H_{M-1}(z^{\frac{1}{M}}W^{M-1})
\end{vmatrix}}{\begin{vmatrix}
H_0(z^{\frac{1}{M}}) & H_0(z^{\frac{1}{M}}W) & \dots & H_0(z^{\frac{1}{M}}W^{M-1}) \\
H_1(z^{\frac{1}{M}}) & H_1(z^{\frac{1}{M}}W) & \dots & H_1(z^{\frac{1}{M}}W^{M-1}) \\
\vdots & \vdots & \ddots & \vdots \\ 
H_{M-1}(z^{\frac{1}{M}}) & H_{M-1}(z^{\frac{1}{M}}W) & \dots & H_{M-1}(z^{\frac{1}{M}}W^{M-1})
\end{vmatrix}} . $}
\end{equation}

\section{Matrix Adaptive Filter}
In the absence of knowledge about the analysis bank, we cannot evaluate the Wiener solution directly. We have to use some iterative method to arrive at it. We use the LMS algorithm~\cite{jn:widlms, bk:haykin} for this purpose. 

A matrix adaptive filter is of this form:
\begin{equation}
\mathbf{a}(n) =  \begin{bmatrix}
\mathbf{a}_{0,0}(n) & \mathbf{a}_{0,1}(n) & \dots & \mathbf{a}_{0,M-1}(n) \\
\mathbf{a}_{1,0}(n) & \mathbf{a}_{1,1}(n) & \dots & \mathbf{a}_{1,M-1}(n) \\
\dots & \dots & \ddots & \dots \\
\mathbf{a}_{M-1,0}(n) & \mathbf{a}_{M-1,1}(n) & \dots & \mathbf{a}_{M-1,M-1}(n) \end{bmatrix}
\end{equation}
where $\mathbf{a}_{p,q}(n)$ is a constituent FIR filter, and it can be represented as
\begin{equation}
\mathbf{a}_{p,q}(n) = [a_{p,q,0}(n),a_{p,q,1}(n),\dots, a_{p,q,L_{p,q}-1}(n)]^T .
\end{equation}
Following similar procedure as given in~\cite{bk:haykin} for scalar adaptive filters, we can derive the LMS algorithm for the matrix adaptive filter which is given by
\begin{align}
\mathbf{a}_{p,q}(n+1) &= \mathbf{a}_{p,q}(n) + \mu_{p,q} e_p(n) \mathbf{v}^*_q(n) , \notag \\
e_p(n) &= d_p(n) - y_p(n) .
\end{align}
The step-size $\mu_{p,q}$ and the filter length $L_{p,q}$ can be chosen by using arguments given in~\cite{bk:haykin,jn:widrow} for convergence. We choose a common filter length and step-size for $\mathbf{a}(n)$ which ensure convergence for each constituent filter.

\section{Results}
\subsection{Experiment 1}
\begin{figure}[!t]
\centering
\resizebox{1\columnwidth}{0.62\columnwidth}{
\includegraphics[width=1\columnwidth, keepaspectratio=true]{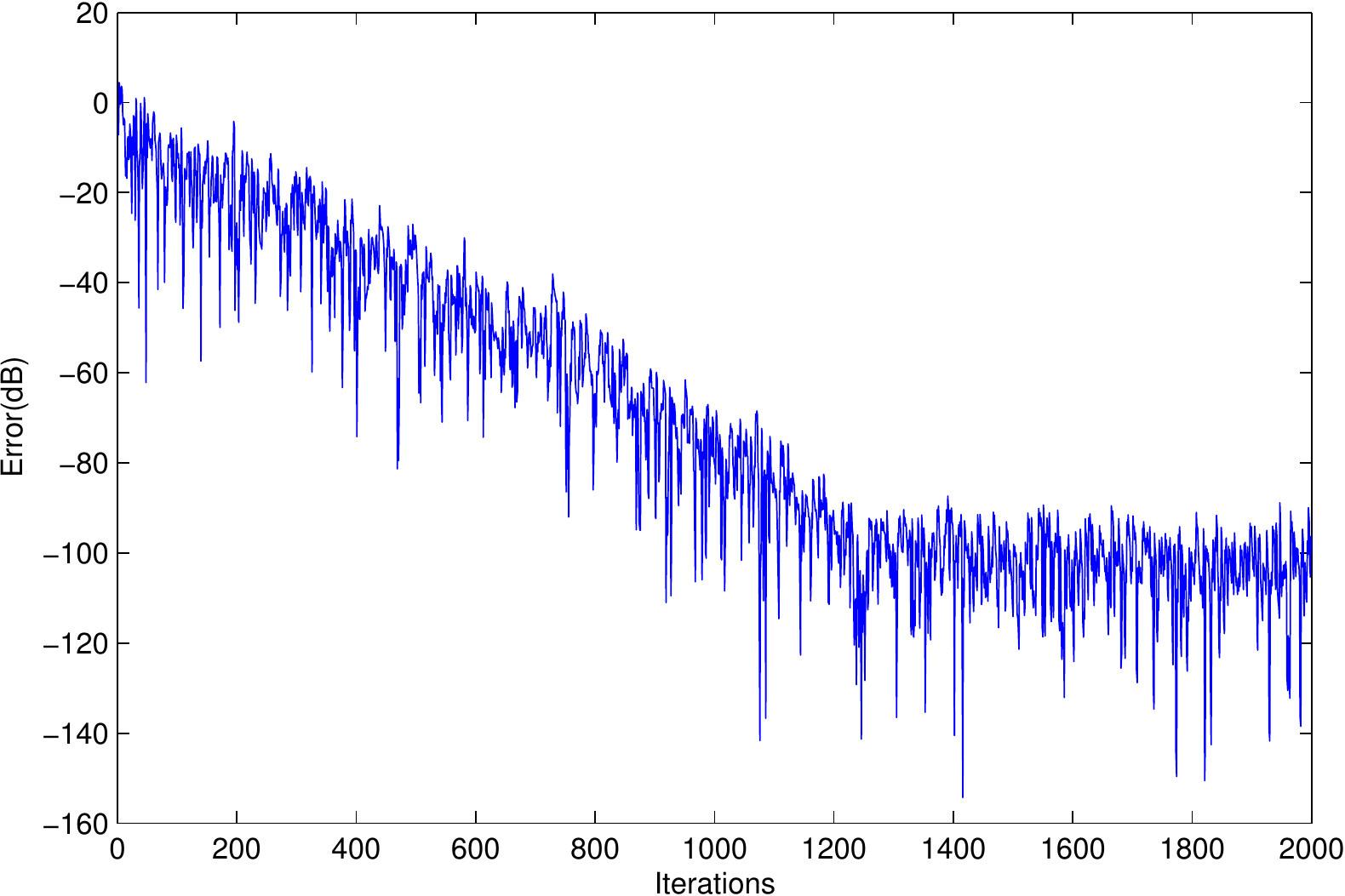}}
\caption{Exp 1: Error vs Iterations}
\label{fig_exp1}
\end{figure}

We considered a 2-band UFB with the following analysis filters
\begin{equation}
\begin{split}
H_0(z) &= 4 + 7z^{-1} + 2z^{-2} , \\
H_1(z) &= 3 -z^{-1} - 1.5z^{-2} .
\end{split}
\end{equation}
A matrix Wiener filter was obtained for $d=0$, which is given by
\begin{equation}
\mathbf{A}(z) = \begin{bmatrix}
\frac{2}{50-17z^{-1}} & \frac{14}{50-17z^{-1}} \\
\frac{3(2-z^{-1})}{50-17z^{-1}} & \frac{-4(2+z^{-1})}{50-17z^{-1}}
\end{bmatrix} .
\end{equation}
The input to the UFB was WSS and it was generated using randn() function of Matlab. For the adaptive filter, we used the Normalized LMS (NLMS) algorithm with the step-size = 0.6 and the filter length = 11. Fig.~\ref{fig_exp1} shows the error vs iteration plot for the adaptive filter. It is clear from Table~\ref{tab_exp1} that the adaptive filter eventually converges to the Wiener filter. 

\begin{table}[!t]
\caption{Exp 1: Adaptive Filter coefficients at iteration = 2000 }
\label{tab_exp1}
\centering
\scalebox{1}{
\begin{tabular}{c c c c }
\midrule
{\centering $a_{1,1}$} & {\centering $a_{1,2}$} & {\centering $a_{2,1}$} & {\centering $a_{2,2}$} \\
\midrule
4.000E-2 & 2.800E-1 &	1.200E-1 &	-1.600E-1 \\
1.360E-2 &	9.520E-2 &	-1.920E-2 & -1.344E-1 \\
4.624E-3 &	3.237E-2 &	-6.528E-3 & -4.570E-2 \\
1.572E-3 &	1.101E-2 &	-2.220E-3 & -1.554E-2 \\
5.350E-4 &	3.741E-3 &	-7.553E-4 & -5.282E-3 \\
1.818E-4 &	1.272E-3 &	-2.566E-4 & -1.796E-3 \\
6.180E-5 &	4.313E-4 &	-8.725E-5 & -6.088E-4 \\
2.087E-5 &	1.455E-4 &	-2.946E-5 & -2.054E-4 \\
6.826E-6 &	4.908E-5 &	-9.636E-6 & -6.929E-5 \\
2.499E-6 &	1.613E-5 &	-3.528E-6 & -2.277E-5 \\
9.423E-7 &	4.791E-6 &	-1.330E-6 & -6.764E-6 \\
\midrule
\end{tabular}
}
\vspace{-13pt}
\end{table}

\subsection{Experiment 2}
\begin{figure}[!t]
\resizebox{1\columnwidth}{0.65\columnwidth}{
\centering
\includegraphics[width=1\columnwidth, keepaspectratio=true]{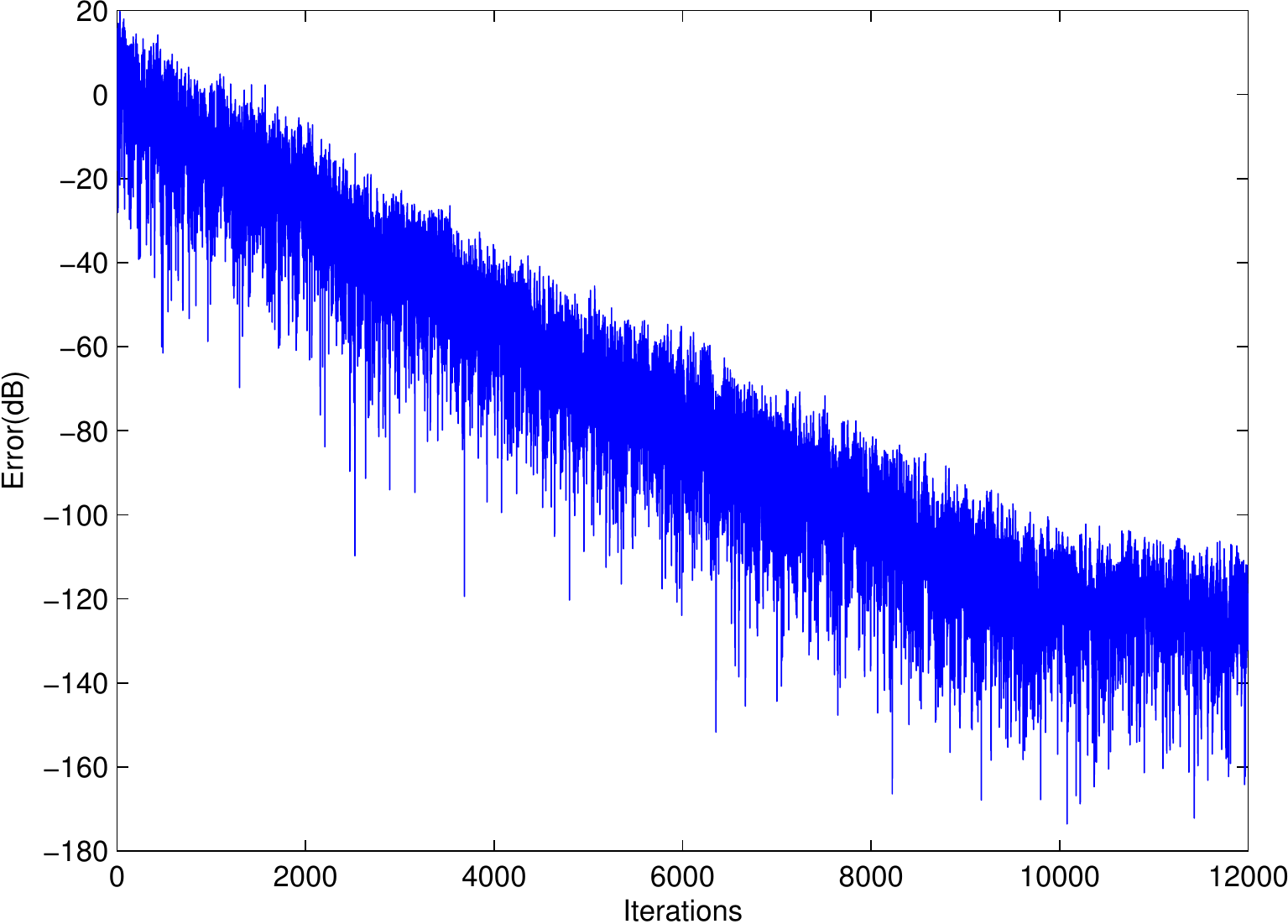}}
\caption{Exp 2: Error vs Iterations}
\label{fig_exp2}
\end{figure}

We next considered a 3-band UFB with
\begin{equation}
\begin{split}
H_0(z) &= 13 - 3z^{-1} + 2z^{-2} - 5z^{-3} - 2z^{-4} , \\
H_1(z) &= 1 -24z^{-1} - 5z^{-2} + 7z^{-3} , \\
H_2(z) &= -19 +5z^{-1} + 14z^{-2} + z^{-3} -8z^{-4} .
\end{split}
\end{equation}
A Wiener filter was obtained for $d=0$, which is given by
\begin{equation}
\arraycolsep=2pt
\medmuskip=1mu
\begin{split}
&\mathbf{A}(z) = (2594 - 642z^{-1} -147z^{-2})^{-1} \times \\
& \resizebox{1\linewidth}{!}{$
\begin{bmatrix}
155.5+20z^{-1} & -26-6z^{-1} & -31.5-5z^{-1} \\
-40.5+51.5z^{-1} & -110+36z^{-1} & -33.5+5.5z^{-1} \\
225.5-25.5z^{-1}+28z^{-2} & 4-82z^{-1}+21z^{-2} & 154.5-71.5z^{-1}-7z^{-2}
\end{bmatrix} . $}
\end{split}
\end{equation}
This Wiener filter is stable. Fig.~\ref{fig_exp2} shows the error vs iteration plot for the adaptive filter where we used the NLMS algorithm with the step-size = 0.45 and the filter length = 15. Table~\ref{tab_exp2} gives the adaptive filter coefficients at iteration = 12000. 
\begin{table}[!t]
\caption{Exp 2: Adaptive Filter coefficients at iteration = 12000 }
\label{tab_exp2}
\centering
\resizebox{1\columnwidth}{!}{
\tabcolsep = 3pt
\begin{tabular}{ccccccccc }
\toprule 
{\centering $a_{1,1}$} & {\centering $a_{1,2}$} & {\centering $a_{1,3}$} & {\centering $a_{2,1}$} & {\centering $a_{2,2}$} & {\centering $a_{2,3}$} & {\centering $a_{3,1}$} & {\centering $a_{3,2}$} & {\centering $a_{3,3}$} \\
\midrule
5.99E-2 & -1.00E-2 & -1.21E-2 &	-1.56E-2 & -4.24E-2 & -1.29E-2 & 8.69E-2 & 1.54E-3 & 5.96E-2 \\
2.25E-2 & -4.79E-3 & -4.93E-3 &	1.60E-2 & 3.38E-3 &	-1.08E-3 & 1.17E-2 & -3.12E-2 & -1.28E-2 \\
8.98E-3 & -1.75E-3 & -1.91E-3 &	3.07E-3 & -1.57E-3 & -9.98E-4 &	1.86E-2 & 4.54E-4 &	-2.50E-3 \\
3.50E-3 & -7.06E-4 & -7.52E-4 &	1.67E-3 & -1.96E-4 & -3.08E-4 &	5.27E-3 & -1.66E-3 & -1.34E-3 \\
1.37E-3 & -2.74E-4 & -2.94E-4 &	5.87E-4 & -1.37E-4 & -1.33E-4 &	2.36E-3 & -3.85E-4 & -4.74E-4 \\
5.39E-4 & -1.08E-4 & -1.15E-4 &	2.40E-4 & -4.51E-5 & -5.03E-5 &	8.82E-4 & -1.89E-4 & -1.94E-4 \\
2.11E-4 & -4.22E-5 & -4.53E-5 &	9.25E-5 & -1.89E-5 & -2.00E-5 &	3.52E-4 & -6.86E-5 & -7.48E-5 \\
8.28E-5 & -1.66E-5 & -1.77E-5 &	3.65E-5 & -7.24E-6 & -7.80E-6 &	1.37E-4 & -2.77E-5 & -2.95E-5 \\
3.24E-5 & -6.50E-6 & -6.96E-6 &	1.43E-5 & -2.87E-6 & -3.06E-6 &	5.39E-5 & -1.08E-5 & -1.15E-5 \\
1.27E-5 & -2.54E-6 & -2.73E-6 &	5.60E-6 & -1.12E-6 & -1.20E-6 &	2.11E-5 & -4.22E-6 & -4.54E-6 \\
4.98E-6 & -9.88E-7 & -1.07E-6 &	2.19E-6 & -4.35E-7 & -4.71E-7 &	8.26E-6 & -1.64E-6 & -1.78E-6 \\
1.94E-6 & -3.90E-7 & -4.15E-7 &	8.55E-7 & -1.71E-7 & -1.83E-7 &	3.22E-6 & -6.47E-7 & -6.88E-7 \\
7.59E-7 & -1.45E-7 & -1.62E-7 &	3.34E-7 & -6.37E-8 & -7.09E-8 &	1.26E-6 & -2.40E-7 & -2.69E-7 \\
2.85E-7 & -5.85E-8 & -6.81E-8 &	1.25E-7 & -2.58E-8 & -3.03E-8 &	4.73E-7 & -9.66E-8 & -1.12E-7 \\
8.30E-8 & -1.08E-8 & -2.35E-8 &	3.66E-8 & -4.50E-9 & -1.00E-8 &	1.38E-7 & -1.85E-8 & -3.99E-8 \\
\bottomrule
\end{tabular}}
\vspace{-13pt}
\end{table}

\section{Conclusion}
In this paper, we have studied the reconstruction characteristics of a UFB using an adaptive filter as the synthesis stage. We first derive the matrix Wiener filter for the UFB, and then we obtain a simplified expression for it. This expression is simple enough to draw useful conclusion about convergence of the matrix adaptive filter. Through experimental results, we have shown the behavior of the adaptive filter. Though we used the NLMS algorithm for the adaptive filter, other variants of the LMS algorithm or other algorithms~\cite{bk:haykin} can also be experimented with.    



\bibliographystyle{IEEEtran}
\bibliography{ufb_ncc}

\end{document}